\documentclass[12pt,preprint]{aastex}
\begin{document}

\title{Formation of X-Ray Cavities by the Magnetically Dominated Jet-Lobe
System in a Galaxy Cluster}


\author{H. Xu\altaffilmark{1,2}, H. Li\altaffilmark{2}, D.
  Collins\altaffilmark{1}, S. Li\altaffilmark{2}, M.L.
  Norman\altaffilmark{1} }

\altaffiltext{1}{Center for Astrophysics and Space Science, University of California, San Diego, 9500
  Gilman Drive, La Jolla, CA 92093.} \altaffiltext{2}{Theoretical
  division, Los Alamos National Laboratory, Los Alamos, NM 87545.}

\begin{abstract}

  We present cosmological magnetohydrodynamic simulations of the
  formation of a galaxy cluster with magnetic energy feedback from an
  active galactic nuclei (AGN). We demonstrate that X-ray cavities can
  be produced by the magnetically dominated jet-lobe system that is
  supported by a central axial current. The cavities are magnetically
  dominated and their morphology is determined jointedly by the
  magnetic fields and the background cluster pressure profile.  The
  expansion and motion of the cavities are driven initially by the
  Lorentz force of the magnetic fields, and the cavities only become buoyant at late
  stages ($> 500$ Myr). We find that up to $80\%-90\%$ of the injected
  magnetic energy goes into doing work against the hot cluster medium,
  heating it, and lifting it in the cluster potential.

\end{abstract}


\section{Introduction}

The absence of spectral signatures of cooling plasmas at the galaxy
cluster centers (Tamura et al. 2001; Peterson et al. 2003) has led to
the suggestion that the intra-cluster medium (ICM) in cluster centers
must be heated.  Powerful radio jet-lobes emanating from supermassive
black holes (SMBHs) in AGNs of clusters are considered to be the
promising heating sources (Binney \& Tabor 1995; Tucker \& David 1997;
see also McNamara \& Nulsen 2007 for a recent review).  High
resolution X-ray images of galaxy clusters by {\it Chandra} have
revealed giant cavities and weak shock fronts in the hot gas (Fabian
et al. 2000; McNamara et al. 2000, 2005), which are commonly
associated with energetic radio lobes (Blanton et al. 2005; Nulsen et
al. 2002) and suggest that magnetic fields play an important role.

Large uncertainties concerning the nature of these cavities,
their formation, evolution, and survivability in the ICM still remain.  Numerical
simulations of hot, underdense bubbles in galaxy clusters have been
performed by a number of authors (e.g., Churazov et al.  2001;
Reynolds, Heinz, \& Begelman 2001; Br\"{u}ggen \& Kaiser 2002; Omma et
al. 2004).  It is generally possible to inject a large
amount of energy into the ICM via AGNs but it is not exactly clear how
the AGN energy can be efficiently utilized (Vernaleo \& Reynolds 2006;
though see Heinz et al.  2006).  One of the most interesting
characteristics of the radio bubbles is that they are intact, whereas
most hydrodynamic simulations \citep{Quilis01,Bruggen02,Dalla04} have
shown that purely hydrodynamic bubbles will disintegrate in timescales
much less than $10^8$ yrs, markedly different from observations.  (we
will use X-ray cavities, bubbles, and radio bubbles interchangeably in
this {\it Letter}).  The stabilizing role of magnetic fields has been
suggested and studied by a few authors (e.g., Jones \& De Young 2005).

A rather different class of models has been proposed and studied, in
which the AGN energy output is modeled in the magnetically dominated
limit (Li et al. 2006; Nakamura, Li, \& Li 2006; see also the previous
work of Blandford 1976, Lovelace 1976, Lynden-Bell 1996, Li et al. 2001).
The key feature of this model is to inject simultaneously both the
poloidal and the (more dominant) toroidal magnetic fields in a small
volume. This is to mimic the possible outcome of an accretion disk
dynamo around an SMBH that shears and twists up the poloidal magnetic
fields and generates large amounts of toroidal fields with an axial
current (as high as $10^{19}$ amperes) flowing along the central axis
of this magnetic structure.  The injection of magnetic fields and
their associated currents lasts a finite time (mimicking the lifetime
of an AGN), after which the magnetic fields and their currents will no
longer be injected but will continue to evolve and gradually dissipate
away. This global current is essential to maintaining the magnetic
structure throughout the lifetime of a magnetic field and its associated
current. The system is not force-free
initially, and so the fields will self-collimate and expand
predominately axially, producing a collimated structure reminiscent of
a ``magnetic tower''.  Extensive 3-D magnetohydrodynamic (MHD)
simulations based on this model in a static cluster-like background, have
demonstrated that such magnetically dominated structures can reproduce
some of the global features of the jet-lobe systems, especially in
maintaining the integrity of the bubbles.

In this {\it Letter}, we present cosmological MHD simulations of a
cluster formation with the feedback of an AGN, with the aim of
understanding the X-ray cavity formation using the magnetically
dominated models proposed by Li et al. (2006). This differs from all
previous studies \citep{Bruggen02,Heinz06} in that the formation of
the X-ray cavity is studied in a realistic and self-consistent
cosmological setting where the dark matter, baryon dynamics and
magnetic fields are all evolved self-consistently. In \S
\ref{sec:method}, we describe our approach and the parameters of the
simulations. In \S \ref{sec:result}, we present the key results.
Conclusions and discussions are given in \S \ref{sec:dis}.

\section{Simulations}
\label{sec:method}

We use the newly developed ENZO+MHD code to simulate the galaxy
cluster formation with the magnetic feedback of SMBHs. This code is an
Eulerian cosmological MHD code with adaptive mesh refinement
\citep{Xu07}.  The galaxy cluster presented here is taken from the
Simulated Cluster Archive of Laboratory of Computational Astrophysics
at University of California, San Diego
(http://lca.ucsd.edu/data/sca/).  It uses a $\Lambda$CDM model with
parameters $h=0.7$, $\Omega_{m}=0.3$, $\Omega_{b}=0.026$,
$\Omega_{\Lambda}=0.7$, and $\sigma_{8}=0.928$. While these are not
precisely the values determined by the latest observations
\citep{Spergel07}, the differences are unimportant for the conclusions
of this study.  For simplicity, the simulation is adiabatic without
the additional physics such as radiative cooling and star formation
feedback. (These effects are not essential for the purpose of this
study.)  The survey volume is 256 $h^{-1}$ Mpc on a side.  (For
simplicity, all spatial scales in the rest of the {\it Letter} are
normalized by the factor $h^{-1}$ and are in comoving unit.)  The
simulations were computed from cosmological initial conditions sampled
onto a $128^3$ root grid and 2 level nested static grids in the
Lagrangian region where the cluster forms which gives an effective
root grid resolution of $512^3$ cells (0.5 Mpc) and dark matter
particles of mass $1.49 \times 10^{10}M_{\odot}$.  Adaptive mesh
refinement (AMR) is used only in the region where the galaxy cluster
forms, with a total of 8 levels of refinement beyond the root grid,
for a maximum spatial resolution of $7.8125$ kpc.  At the center of
the cluster when the AGN is ``turned on'' (see below), we further
increased the resolution at the injection region to be $\sim 1.95$
kpc.  While the baryons are resolved at higher and higher spatial and
mass resolution at higher levels, the dark matter particles maintain
constant in mass so as not to add any additional noise. The
simulations were evolved from redshift $z=30$ to $z=0$.

The energy output from an SMBH is simulated by injecting magnetic
energy in a small volume centered on a chosen massive galaxy that is
at or near the center of the cluster. It is currently not possible to
resolve both the galaxy cluster and the black hole environment
simultaneously, so we have adopted an approach that mimics the
possible magnetic energy injection by an SMBH \citep{Li06}.  We start
the magnetic energy injection at redshift $z=0.05$, at which the
cluster has a virial radius $r_v \approx 1.5$ Mpc, a virial mass $m_v
\approx 10^{15} M_{\odot}$, baryon to dark matter ratio $8.16\%$. In
the cluster center where the chosen AGN is located, the gas density is
$7.27\times 10^{-27}$ g cm$^{-3}$, pressure $1.49 \times 10^{-10}$ dyn
cm$^{-2}$, and temperature $1.45 \times 10^{8} K$.  The density
profile of the ICM can be fitted as: $\rho = \rho_0
[1+(\frac{r}{r_c})^2]^{-3k/2}$ with $r_c = 100$ kpc and $k = 0.485$.
Magnetic fields are injected for 36 Myr in a region of $\sim 14$ kpc
with a peak strength of $\sim 200 \mu G$. There are about $10^{9}\
M_{odot}$ mass enclosed in the injection region.  The evolution from
$z=0.05$ to $z=0$ lasts about $670$ Myrs.  So the cluster is still
evolving while the magnetic fields from an AGN are injected and are
piercing through the ICM.  Many different simulations were performed
with a range of injection parameters, but we present the results
mostly using one simulation, while the conclusions drawn are general.

Note that after the first 36 Myr, magnetic fields and their associated
currents are no longer injected. This, however, does not mean that the
magnetic fields (and their currents) will disappear immediately.
Instead, they will continue to evolve, gradually losing energy
(especially in the early stages) by, say, doing work against the
background ICM. Since the typical dynamic time is $\sim 10^8$ yrs and
our simulation lasts $7\times 10^8$ yrs, we do not expect the
numerical dissipation of magnetic energy to play a major role.  We
have used high resolution on all the regions with magnetic fields (via
AMR), so the numerical dissipation timescale should be much longer
than the dynamic timescale.

\section{Results}
\label{sec:result}

\subsection{Stages of X-ray Cavity Formation}

\begin{figure}
\begin{center}
\includegraphics[width=\textwidth]{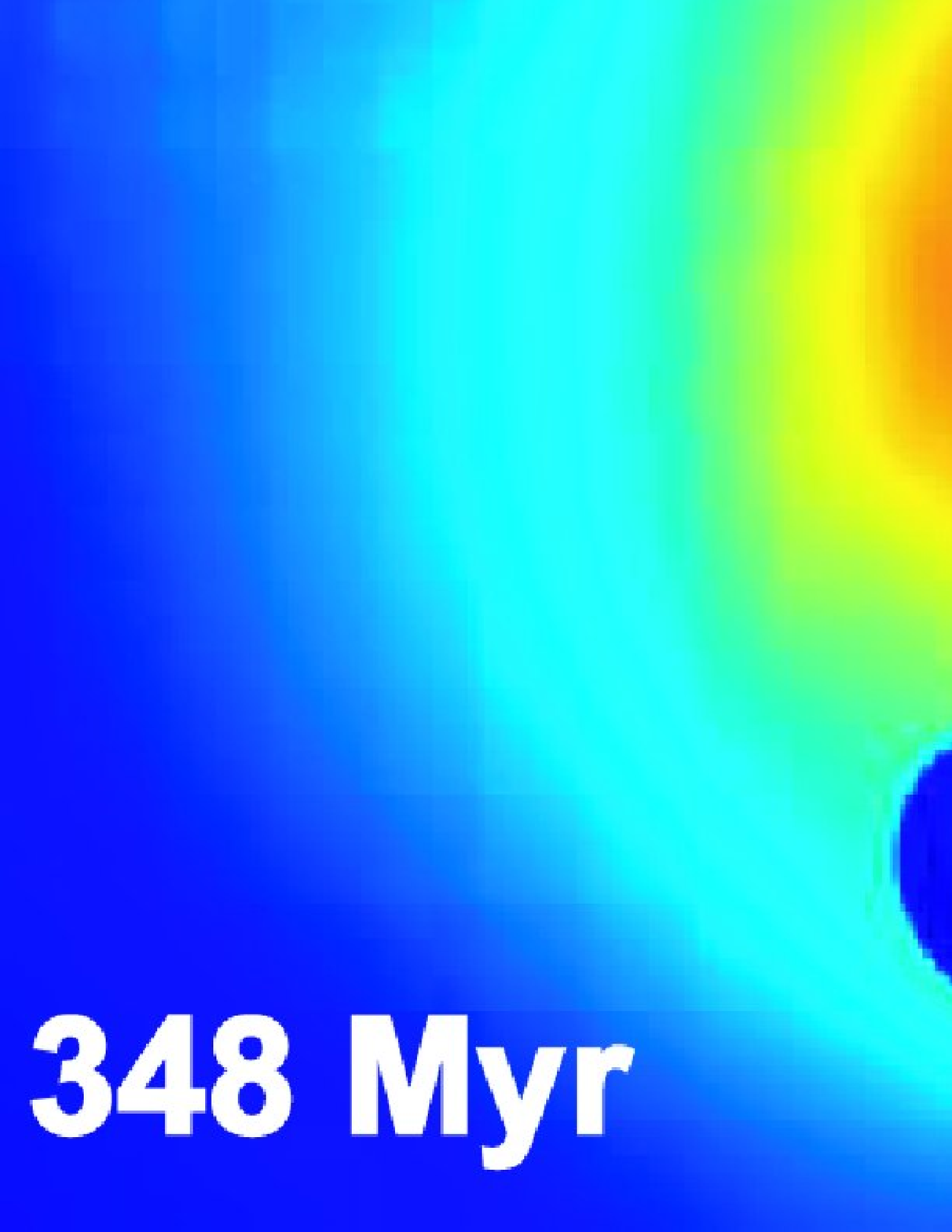}
\end{center}
\caption{Snapshots of the jet-lobe evolution driven 
  by the magnetic energy output of an AGN. Each image is
  672 kpc on the side. Columns from left to right are slices of
  density, temperature, the averaged magnetic energy density and the
  integrated X-ray luminosity, respectively. The top row
  shows the cluster with the jet-lobe at the end of magnetic energy 
  injection.  The middle and bottom rows show the
  well-developed bubbles moving out of the cluster center. The bubbles
  are driven by magnetic forces at all stages, and might become
  buoyant only after $t > 500$ Myr.  (Each image uses its own color
  map as described in text.)
  \label{fig:evolution}}
\end{figure}

Images of the cluster at different epochs are given in Fig.
\ref{fig:evolution}.  The ranges of density are, from top to bottom,
 $3.70 \times 10^{-28} - 8.77 \times 10^{-27}$, $9.41 \times 10^{-29} - 6.24\times 10^{-27}$,
  and $1.16 \times 10^{-28} - 5.67 \times 10^{-27}$ g cm$^{-3}$,
respectively. A cutoff of $2\times 10^8$ K is used in all temperature
images. The minimum temperatures are $4.65\times 10^7$, $6.78\times
10^7$ and $6.9\times 10^{7}$ K from top to bottom. The maximum
magnetic energy density of each image is $4.16\times 10^{-11}$, $1.67\times 10^{-11}$, and $6.05\times
10^{-12}$ erg cm$^{-3}$. The ranges of path-integrated X-ray emission are
$6.67\times 10^{-7} - 5.26\times 10^{-5}$, $6.16\times
10^{-7} - 4.49\times 10^{-5}$ and $4.25\times 10^{-7} -
4.41\times 10^{-5}$ erg cm$^{-2}$ s$^{-1}$, from top to bottom, respectively.
The integrated X-ray intensity is taken from the $0.5-7.5$ keV band
with $0.3$ solar metal abundance assumed. These projected results are
obtained by integrating $336$ kpc along the line of sight.

\begin{figure}
\includegraphics[width=\textwidth]{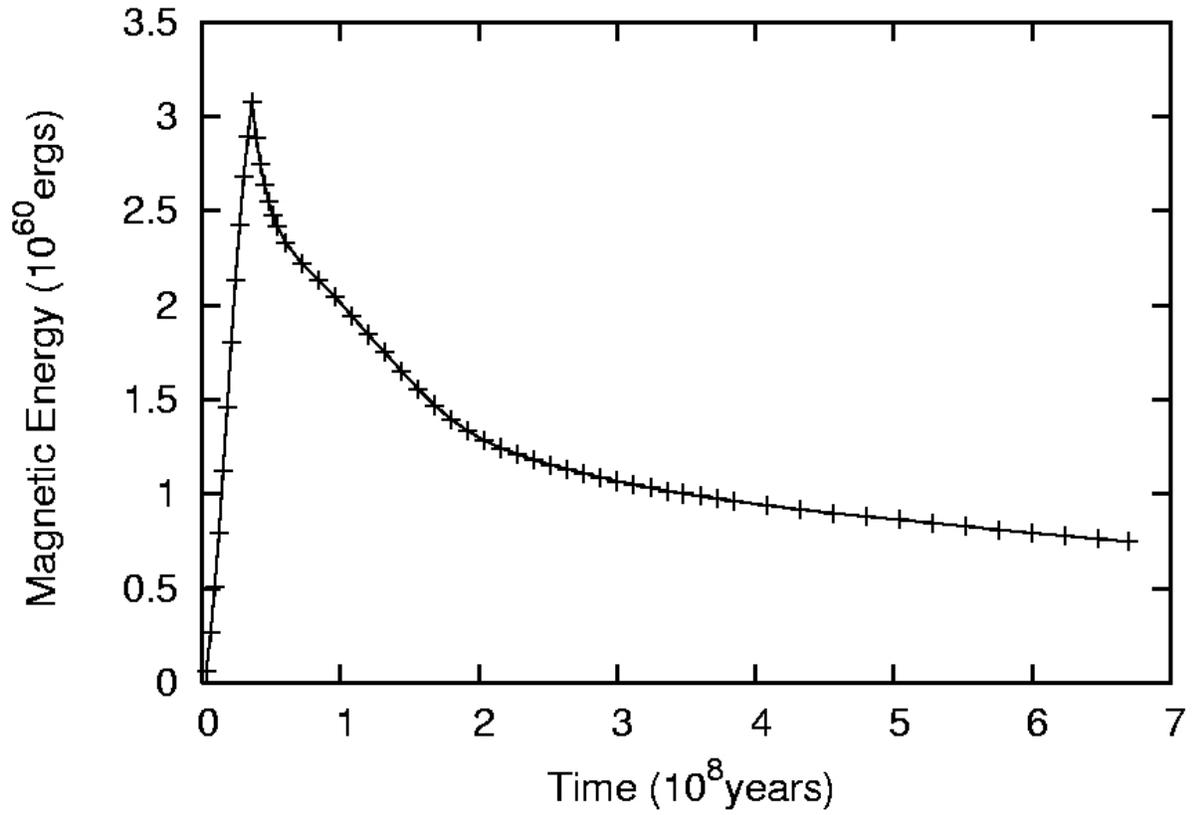}
\caption{Evolution of the total magnetic field energy in the
  cluster from redshift $z=0.05$ to $z=0$.  The initial increase comes
  from the injection (which ends at $36$ Myr), followed by a continual
  decline as the jet-lobe system converts magnetic energy to
  thermal, kinetic and gravitational energies of the background ICM.
  \label{fig:EB}}
\end{figure}

A pair of cavities can be seen in the images at late times, which are
created by the expansion of magnetic fields as they move out of the
cluster core region. Note the relative motion between the cluster core
and the injection location. The evolution of the total magnetic energy
(integrated over the whole cluster) is given in Figure \ref{fig:EB}.
We can see that the system first goes through an injection phase from
$t = 0 - 36$ Myr, during which $\sim 6 \times 10^{60}$ ergs of
magnetic energy is injected.  This gives an average input power of
$5.3 \times 10^{45}$ ergs s$^{-1}$, which is comparable to the typical
luminosity of a powerful AGN.  At the end of the injection, only about
$3.1\times 10^{60}$ ergs still remain in the magnetic form while the
rest of the injected magnetic energy has all been transferred to the
surrounding ICM.  At $t=48$ Myr, the poloidal current is $\sim
1.9\times 10^{19}$ amperes along the central axis, whose corresponding
toroidal magnetic fields self-collimate the outflow, which undergoes a
supersonic expansion in the cluster core region \citep{Li06,
  Nakamura06} and forms a ``tower-like'' structure.


Once the injection ceases, the magnetic energy in the cluster starts
to decrease, first at a fast rate ($\sim 10^{52}$ ergs yr$^{-1}$) from $t
\sim 40 - 200$ Myr, followed by a much slower rate ($\sim 10^{51}$
ergs yr$^{-1}$) from $t \sim 200 - 670$ Myr. This transition coincides with
the time when the magnetic lobes leave the cluster core region ($\sim
100$ kpc, the middle row of Fig.  \ref{fig:evolution}). The plasma
density and pressure of the ICM drop rapidly beyond the core, hence
the slower rate of magnetic energy dissipation. The poloidal current
is $\sim 5\times 10^{18}$ ampere at $t=348$ Myr.  Once the magnetic
lobes leave the cluster core, they experience additional expansion,
forming the round or flattened shapes, depending on the surrounding
cluster pressure environment (the bottom row of Fig.
\ref{fig:evolution}).  The final total magnetic energy in the cluster
is about $7.5\times 10^{59}$ ergs, roughly $\sim 10\%$ of the total
injected magnetic energy. This implies that $\sim 90\%$ of the
injected magnetic energy has been transferred to the ICM: their
thermal energy (heating via compression and shocks), their kinetic
energy (driving the bulk flows), and their potential energy (being
lifted in the cluster potential well).

\subsection{Shock Fronts and Cavity Properties}

{\it Chandra} observations have revealed that shocks around the radio
lobes are weak \citep{Fabian03, Nulsen05, McNamara05}.  In our
simulations, a global shock front enclosing the whole magnetic
structure is generated early by the injection of magnetic fields.  At
$t=48$ Myr, the shock's Mach number is about $1.55$ at a distance
roughly $92$ kpc away from the central galaxy.  This shock, which is
hydrodynamic in nature, gradually weakens as it moves ahead of and
away from the magnetic structure.  By $t=168$ Myr, the shock is about
$261$ kpc away and barely visible (the middle row of Fig.
\ref{fig:evolution}), with a Mach number just above $1$.  After this
time, the shock is dissipated into the background ICM.

The Lorentz forces from the magnetic fields expand the magnetic
structure and push the ICM away.  In the nearly ideal MHD limit, the
plasma is ``frozen in'' with the magnetic fields, and the mixing between
the magnetized jet-lobe system and the background ICM is inhibited. So
the plasma density associated with the magnetic structure decreases as
the jet expands, creating the density cavities.  The magnetic fields
undergo significant lateral expansion as they leave the central
core region (the middle and bottom rows of Fig.  \ref{fig:evolution}).
The formation of such large, relatively ``round'' lobes is jointly
determined by the ambient ICM pressure and an axial electric current
which flows along the ``spine'' of the magnetic structure and returns
around the outer boundary of the lobes and jets. The central poloidal
current generates toroidal magnetic fields whose strength behaves as
$1/r$ where $r$ is the cylindrical distance from the jet axis. Such a
decreasing magnetic pressure ($\propto 1/r^2$) is eventually balanced
by the background ICM pressure, thus determining the size of the X-ray
cavity.  More detailed comparisons with a large set of bubble
observations show that such a magnetically dominated model agrees
quite well with bubbles' size and distance distributions
\citep{Diehl08}.

\begin{figure}
\includegraphics[width=\textwidth]{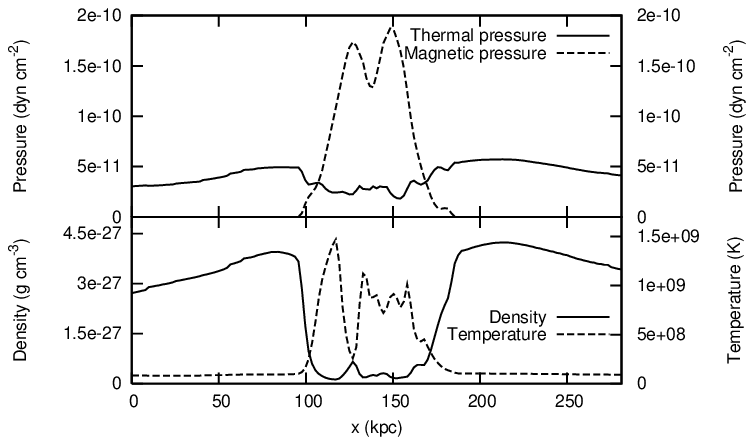}
\includegraphics[width=\textwidth]{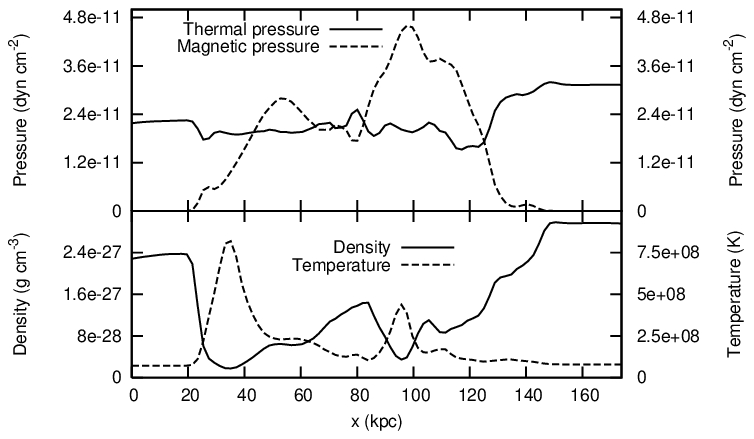}
\caption{Spatial distribution of magnetic pressure, thermal pressure,
  density, and temperature along a horizontal line through
  the southern lobe at $t=168$ Myr ({\it top}) and $t=348$ Myr ({\it
    bottom}).  The lobe is dominated by the magnetic
  pressure. 
  \label{fig:cut}}
\end{figure}

To illustrate the internal property of the lobes, we take a horizontal
``cut'' through the southern lobe at $t = 168$ and $348$ Myr (Fig.
\ref{fig:evolution}) and plot its properties along the line in Fig.
\ref{fig:cut}. At $t=168$ Myr, the cavity is clearly dominated by
magnetic energy. The peak magnetic field strength is probably higher 
than the inferred typical field value from observations. 
At $t=368$ Myr, it has a comparable amount of total
magnetic energy ($\sim 2.4\times 10^{59}$ ergs) and thermal energy
($\sim 2.2\times 10^{59}$ ergs). The total kinetic energy is much
smaller, $\sim 3.6\times 10^{58}$ ergs.  The implied enthalpy upper
limit of this lobe is $4pV \approx 8.8\times 10^{59}$ ergs.  Assuming
that the northern lobe has a similar energetics as the southern lobe,
the total enthalpy in the lobes is $\sim 1.8 \times 10^{60}$ ergs,
which is much smaller than the amount of magnetic energy that has been
transferred, $\sim 5.2\times 10^{60}$ ergs.  So the enthalpy could be
a serious under-estimate of the injected magnetic energy. 


It has often been suggested that the evolution of the X-ray bubbles is
driven by buoyancy. Using lobe's size, distance from the galaxy and 
cluster gravity, we have computed the implied
buoyancy time by pretending they were buoyant. We found that the
calculated buoyancy times are $\sim 236$ and $437$ Myr, whereas the
actual evolution times are $168$ and $348$ Myr, respectively.  
The difference does get smaller at late times ($> 500$ Myr).

\section{Conclusions and Discussions}
\label{sec:dis}

With our cosmological MHD simulations, we find that, in the realistic cluster
environment where the ICM plasma interacts dynamically with the
magnetic jet-lobe, X-ray cavities can naturally form using the
magnetically dominated models proposed by Li et al. (2006). The
magnetic fields inside the bubbles stabilize the interface
instabilities so that bubbles can remain intact. The lifetime of these
bubbles can be quite long and only become truly buoyant probably after
$\sim 500$ Myr.

We have performed additional cosmological MHD simulations with
radiative cooling and star formation feedback and found that our
conclusions for the X-ray cavity formation mechanism do not change.
While we have demonstrated the formation of X-ray cavities, much more
studies are needed in order to address comprehensively the cooling
flow problem at the cluster cores.  The present simulation, with just
one AGN, already has important implications for understanding the ICM
heating problem. Up to $80\%-90\%$ of the injected energy has been
dissipated in the surrounding ICM. Further studies are underway with
AGNs at different redshifts that have different cluster environments
so that we can gain a comprehensive understanding of the overall
heating of the ICM by AGNs. These simulations will be presented in
future publications.  The morphology of the jet-lobe is dependent on
the background density radial profile, which is different for massive
clusters (such as the one presented here) and groups or poor clusters.
Future work will address this issue.

\acknowledgments{We thank the referee whose comments have improved the
presentation of this letter. H. Li thanks S. Colgate for discussions. This work
  was supported by the LDRD and IGPP programs at LANL.  ENZO was
  developed at the Laboratory for Computational Astrophysics, UCSD,
  with partial support from NSF grant AST-0708960 to M. L. Norman. }

\clearpage

\begin{thebibliography}{}

\bibitem[Binney \& Tabor (1995)]{Binney95} Binney, J. \& Tabor, G. 1995,
  MNRAS, {276,} 663


\bibitem[Blandford (1976)]{B76}
        Blandford, R. D. 1976, \mnras, 176, 465

\bibitem[Blanton et al. (2005)]{Blanton01} Blanton, E. L., Sarazin, C.
  L., McNamara, B. R., \& Wise, M. W. 2001, ApJ, {558,} L15

\bibitem[Br\"{u}ggen \& Kaiser (2002)]{Bruggen02} Br\"{u}ggen, M. \&
  Kaiser, C. R. 2002, Nature, {418,} 301


\bibitem[Churazov et al. (2001)]{Churazov01} Churazov, E. et al. 2001,
  ApJ, {554,} 261


\bibitem[Dalla Vecchia et al. (2004)]{Dalla04} Dalla Vecchia, C. et al.,
  2004, MNRAS, \textbf{355,} 995

\bibitem[Diehl et al. (2008)]{Diehl08} Diehl, S., Li, H., Fryer, C. \&
  Rafferty, D. 2008, ApJ, submitted (astro-ph/0801.1825)


\bibitem[Fabian et al. (2000)]{Fabian00} Fabian, A. C. et al. 2000,
  MNRAS, {318,} L65

\bibitem[Fabian et al. (2003)]{Fabian03} Fabian, A.C. et al. 2003,
  MNRAS, {344,} L43




\bibitem[Heinz et al. (2006)]{Heinz06} Heinz, S.,Br\"{u}ggen M., Young,
  A.,  \& Levesque,
  E. 2006,MNRAS, {373,} L65

\bibitem[Jones \& De Young (2005)]{Jones05} Jones, T. W. \& De Young, 
D. S. 2005, ApJ, {624,}, 586

\bibitem[Li et al. (2001)]{Li01}
        Li, H., Lovelace, R. V. E., Finn, J. M., \& Colgate,
        S. A. 2001,  \apj, 561, 915

\bibitem[Li et al. (2006)]{Li06} Li, H., Lapenta, G, Finn, J.M., Li, S.
  \& Colgate, S. A. 2006, ApJ, {643,} 92


\bibitem[Lovelace (1976)]{L76}
        Lovelace, R. V. E. 1976, Nature, 262, 649

\bibitem[Lynden-Bell (1996)]{Lynden96} Lynden-Bell, D. 1996, MNRAS, 279, 389

\bibitem[McNamara et al. (2000)]{McNamara00} McNamara, B.R. et al. 2000,
  ApJ,  {534,} L135 

\bibitem[McNamara et al. (2005)]{McNamara05} McNamara, B.R. et al. 2005,
  Nature, {433,} 45

\bibitem[McNamara \& Nulsen (2007)]{McNamara07} McNamara, B.R. \&
  Nulsen, P.E.  2007, ARA\&A, {45,} 117

\bibitem[Nakamura{,} Li, \& Li (2006)]{Nakamura06}  Nakamura, M., Li, H. \&
  Li, S. 2006, ApJ, {652,} 1059


\bibitem[Nulsen et al. (2002)]{Nulsen02} Nulsen, P. E. J. et al. 2002,
  ApJ,  {568,} 163

\bibitem[Nulsen et al. (2005)]{Nulsen05} Nulsen, P.E., McNamara, B.R.,
  Wise M.W.  \& David, L.P. 2005, ApJ, {628,} 629


\bibitem[Omma et al. (2004)]{Omma04} Omma, H. et al. 2004, MNRAS, {348,} 1105


\bibitem[Peterson et al. (2003)]{Peterson03} Peterson, J. R. et
  al. 2003, ApJ,  {590,} 207

\bibitem[Quilis{,} Bower \& Balogh (2001)]{Quilis01} Quilis, V., Bower,
  R. G.  \& Balogh, M.L. 2001, MNRAS, 328, 1091 


\bibitem[Reynolds{,} Heinz, \& Begelman (2001)]{Reynolds01} Reynolds,
  C. S.,  Heinz, S, \& Begelman, M. C. 2001, ApJ, {549,} L179



\bibitem[Spergel et al. (2007)]{Spergel07} Spergel, D. N. et al. 2007,
  ApJS,  {170,} 377

\bibitem[Tamura et al. (2001)]{Tamura01} Tamura, T. et al. 2001, A\&A,  
{365,} L87

\bibitem[Tucker \& David (1997)]{Tucker97} Tucker, W. \& David,
  L. P. 1997,   ApJ, {484,} 602

\bibitem[Vernaleo \& Reynolds (2006)]{Vernaleo06} Vernaleo, J. C. \&
  Reynolds, C. S. 2006, ApJ, {645,} 83


\bibitem[Xu et al. (2008)]{Xu07} Xu, H. et al. 2008, in AIP Conf. Proc. 990, 
  First Stars III, eds B.W. O'Shea, A.  Heger \& T. Abel (Melville: AIP), 36
        

\end{thebibliography}
\end{document}